\documentclass[reprint,superscriptaddress,amsmath,amssymb,prl]{revtex4-1}

\usepackage{graphicx}
\usepackage{dcolumn}
\usepackage{bm}
\usepackage{gensymb}

\newcommand{\srs}{Sr$_3$Rh$_4$Sn$_{13}$}
\newcommand{\csrs}{(Ca$_x$Sr$_{1-x}$)$_3$Rh$_4$Sn$_{13}$}
\newcommand{\csis}{(Ca$_x$Sr$_{1-x}$)$_3$Ir$_4$Sn$_{13}$}

\begin{document}

\title{Order-Disorder Transitions in \csrs}

\author{Puspa Upreti}
\affiliation{Materials Science Division, Argonne National Laboratory}
\affiliation{Department of Physics, Northern Illinois University}
\author{Matthew Krogstad}
\affiliation{Materials Science Division, Argonne National Laboratory}
\author{Charlotte Haley}
\author{Mihai Anitescu}
\author{Vishwas Rao}
\affiliation{Mathematics and Computer Science Division, Argonne National Laboratory}
\author{Lekh Poudel}
\affiliation{Department of Materials Science and Engineering, University of Maryland}
\affiliation{NIST Center for Neutron Research, National Institute of Standards and Technology}
\author{Omar Chmaissem}
\affiliation{Materials Science Division, Argonne National Laboratory}
\affiliation{Department of Physics, Northern Illinois University}
\author{Stephan Rosenkranz}
\author{Raymond Osborn}
 \email{rosborn@anl.gov}
\affiliation{Materials Science Division, Argonne National Laboratory}

\date{\today}

\begin{abstract}
The classification of structural phase transitions as displacive or order-disorder in character is usually based on spectroscopic data above the transition. We use single crystal x-ray diffraction to investigate structural correlations in the quasi-skutterudites, \csrs, which have a quantum phase transition at $x\sim0.9$. Three-dimensional pair distribution functions show that the amplitudes of local atomic displacements are temperature-independent below the transition and persist to well above the transition, a signature of order-disorder behavior. The implications for the associated electronic transitions are discussed. 

\end{abstract}


\maketitle

For over 60 years, structural phase transitions have been discussed in terms of two categories that represent different limits of the anharmonic potentials driving the distortions \cite{Cochran:1960hu, Bruce:1980bv, Onodera:2004, Bussmann-Holder.2006}. In the first category, displacive transitions result from the condensation of soft phonon modes at the wavevector of superlattice peaks in the low-temperature phase \cite{Cochran:1960hu}. The second category, order-disorder transitions, requires that the distorted atoms occupy sites located at the minima of a deep multiwell potential. These sites are randomly occupied at high temperature, but as the temperature is lowered, site occupations in neighboring unit cells become increasingly  correlated, and the transition occurs when long-range phase coherence develops. In the extreme order-disorder limit, the quasi-harmonic phonon spectra above the structural transition temperature, T$_s$, are not renormalized, but a quasi-elastic peak, usually called a ``central'' peak, is observed from tunneling between the potential wells \cite{Bruce:1980bv, Onodera:2004}. 

It is now commonly believed that few materials rigidly belong to either category but instead fall on a broad spectrum between the two limits \cite{Bussmann-Holder.2006, Armstrong.1989}. Such classifications have usually depended on spectroscopic properties measured above the transition, such as one-phonon spectra accessible through inelastic neutron or x-ray scattering \cite{Bruce:1980bv}. However, the data are often ambiguous; materials that are considered to undergo displacive transitions often display a central peak due to anharmonicity or precursor critical fluctuations \cite{Bruce:1980bv}, while there can be soft mode behavior in systems that display the characteristics of an order-disorder transition, such as local distortions persisting well above T$_s$ \cite{Stern:1996}. 

\begin{figure}[!b]
\includegraphics[width=\columnwidth]{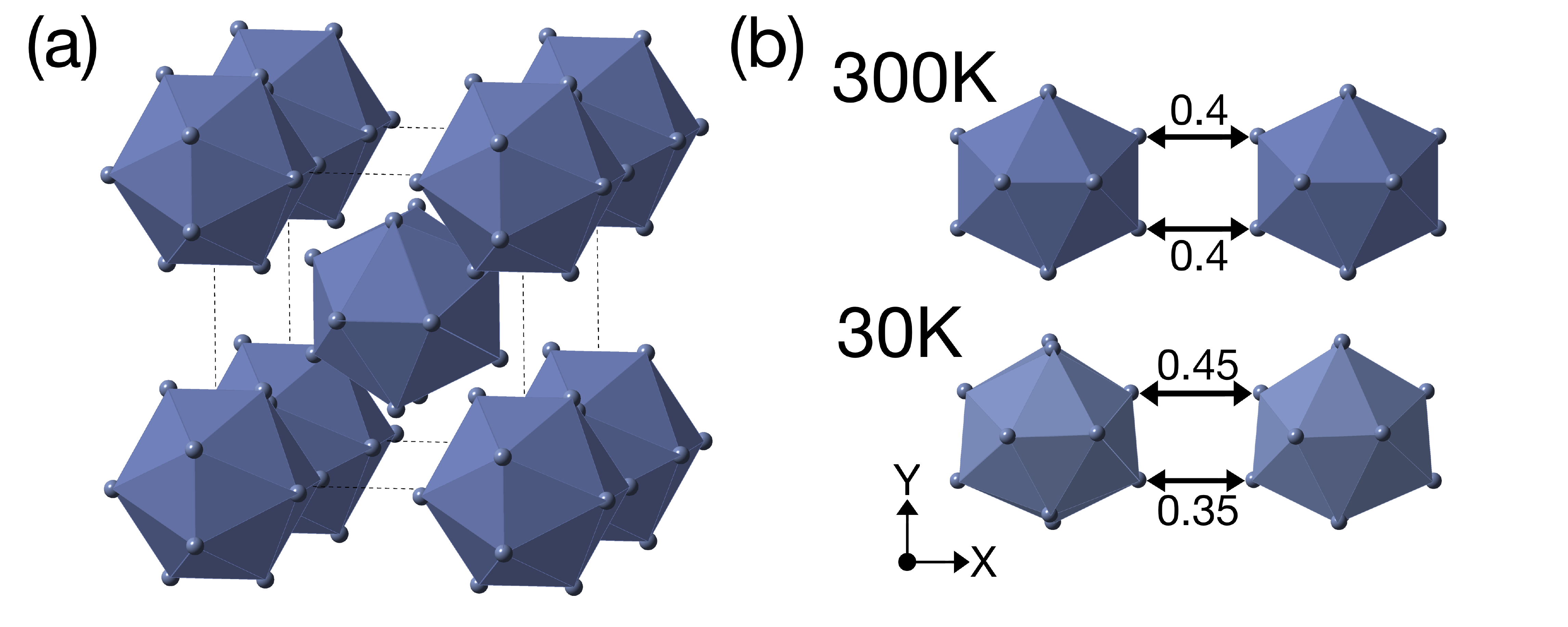} 
\caption{\label{fig:crystal} (a) The unit cell of  the high temperature cubic phase of \csrs\ with only the tin sites shown. The Sn(1) and Sn(2) sites are at the centers and vertices of the icosahedra, respectively. All the Sn(1)-Sn(2) bond lengths are identical above the structural phase transition. (b) Two neighboring icosahedra viewed along the $z$-axis, showing an example of the atomic distortions along the X-axis below the transition.}
\end{figure}

\begin{figure*}[!t]
\includegraphics[width=2\columnwidth]{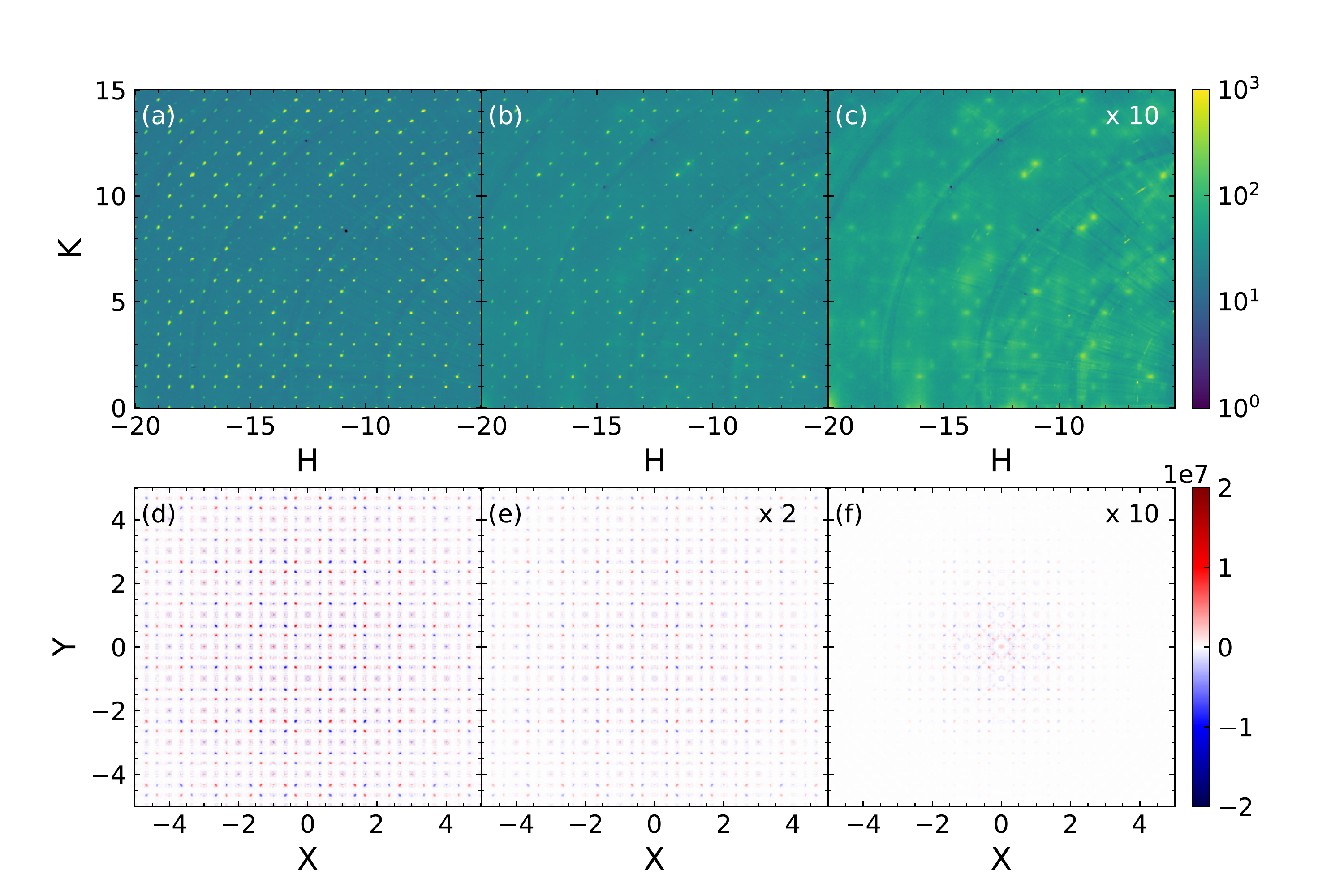} 
\caption{\label{fig:data} (a-c) S(\textbf{Q}) of \srs\ measured in the (HK$\frac{1}{2}$) plane at (a) 30~K. (b) 120~K, and (c) 150~K. (d-f) 3D-PDF in the (XY$\frac{1}{2}$) plane at (d) 30~K, (e) 120~K, and (f) 150~K. The axes are in units of (a-c) 2$\pi/a$ and (d-f) $a$, respectively ($a=9.8$~\AA). }
\end{figure*}

The problem with basing the classification on measurements performed above the phase transition is that thermally activated disorder blurs the distinction between the two categories, particularly when the temperature is comparable to any potential barriers. Furthermore, such experiments do not address the most salient characteristic of order-disorder transitions, namely that, within the ordered phase, the amplitudes of local distortions are to a first approximation fixed by the positions of the potential well minima. In an order-disorder transition, the temperature dependence of the order parameter, observed as an increase in the superlattice peak intensities below T$_s$, is due to changes in the site occupations, not changes in the distortion amplitude. To distinguish between these two, one must also measure fluctuations in the local order parameter in addition to the average value. Recent advances in x-ray instrumentation now allow large contiguous volumes of reciprocal space to be measured rapidly in single crystals, thus enabling both Bragg peaks from the average structure and diffuse scattering from local disorder to be transformed into three-dimensional pair-distribution-functions (PDF). These provide real space maps of interatomic vector probabilities \cite{Weber:2012en, Krogstad:2019tc} that contain direct, model-independent measures of displacements in the low temperature phase, allowing both the local distortions and their correlations up to 200~\AA\ or more to be determined \cite{Krogstad:2019tc}.

The reason why the classification matters is that it is key both to understanding both the origin of the structural phase transition and the nature of the fluctuations in the disordered phase. In this letter, we show how it helps to explain the interplay between structural and electronic fluctuations in the quasi-skutterudite compounds, \csrs. This family of compounds undergoes second-order structural phase transitions from a simple cubic structure (\textit{Pm$\overline{3}$n}), usually labeled the \textit{I} phase \cite{Miraglia.1986,Goh:2015cl}, to a body-centered cubic \textit{$I^\prime$} superstructure (\textit{I$\overline{4}$3d}). Both contain icosahedra of tin atoms (Fig. \ref{fig:crystal}), with the Sn(1) site at their center and 12 Sn(2) sites at their vertices \cite{Miraglia.1986}. The transition involves displacements of the Sn(2) sites that result in a doubling of the unit cell lattice parameter. In the related family, \csis, Klintberg \textit{et al} proposed that the phase transition is due to a charge density wave (CDW) instability \cite{Klintberg:2012dt}, an interpretation apparently supported by shifts in spectral weight in optical spectroscopy measurements observed in both sets of compounds \cite{Fang:2014kr, Ban.2017},  although the nature of the charge disproportionation at T$_s$ has never been established.

These compounds are also superconducting \cite{Kase.2011}, with transition temperature that peak when the structural phase transition is suppressed by either applied pressure or, equivalently, chemical pressure by calcium substitution \cite{Klintberg:2012dt, Goh:2015cl}. In \csrs, T$_s\sim$135~K and the superconducting transition temperature, T$_c = 4.7$~K, at $x=0$, but T$_s$ falls to 0 at $x\approx 0.9$, where T$_c = 7.8$~K. This makes it reminiscent of quantum phase transitions observed in other doped CDW systems \cite{Morosan:2006hk}, implying a phase competition between CDW and superconducting order.

We report a 3D PDF analysis of \csrs, with $x=0$, 0.1, 0.6, and 0.65, using the methods described in Ref. \citenum{Krogstad:2019tc}, which shows that the PDF maps reveal the magnitude of the local tin atom displacements in the \textit{$I^\prime$} phase. Our main observation is that, at all $x$, the displacement amplitude is independent of temperature below T$_s$, and even above T$_s$ (up to at least 200~K at $x=0$) where the structural correlations are short-range. Our results therefore show unambiguously that T$_s$ marks an order-disorder transition, \textit{i.e.}, the temperature at which the phase coherence of the distorted-site occupations becomes long-range, rather than the temperature at which the static distortions collapse. This has important implications for the nature of the electronic reconstruction at T$_s$ and the electronic fluctuations above T$_s$ close to the quantum critical point, which we discuss at the end.

\begin{figure}[!b]
\includegraphics[width=0.8\columnwidth]{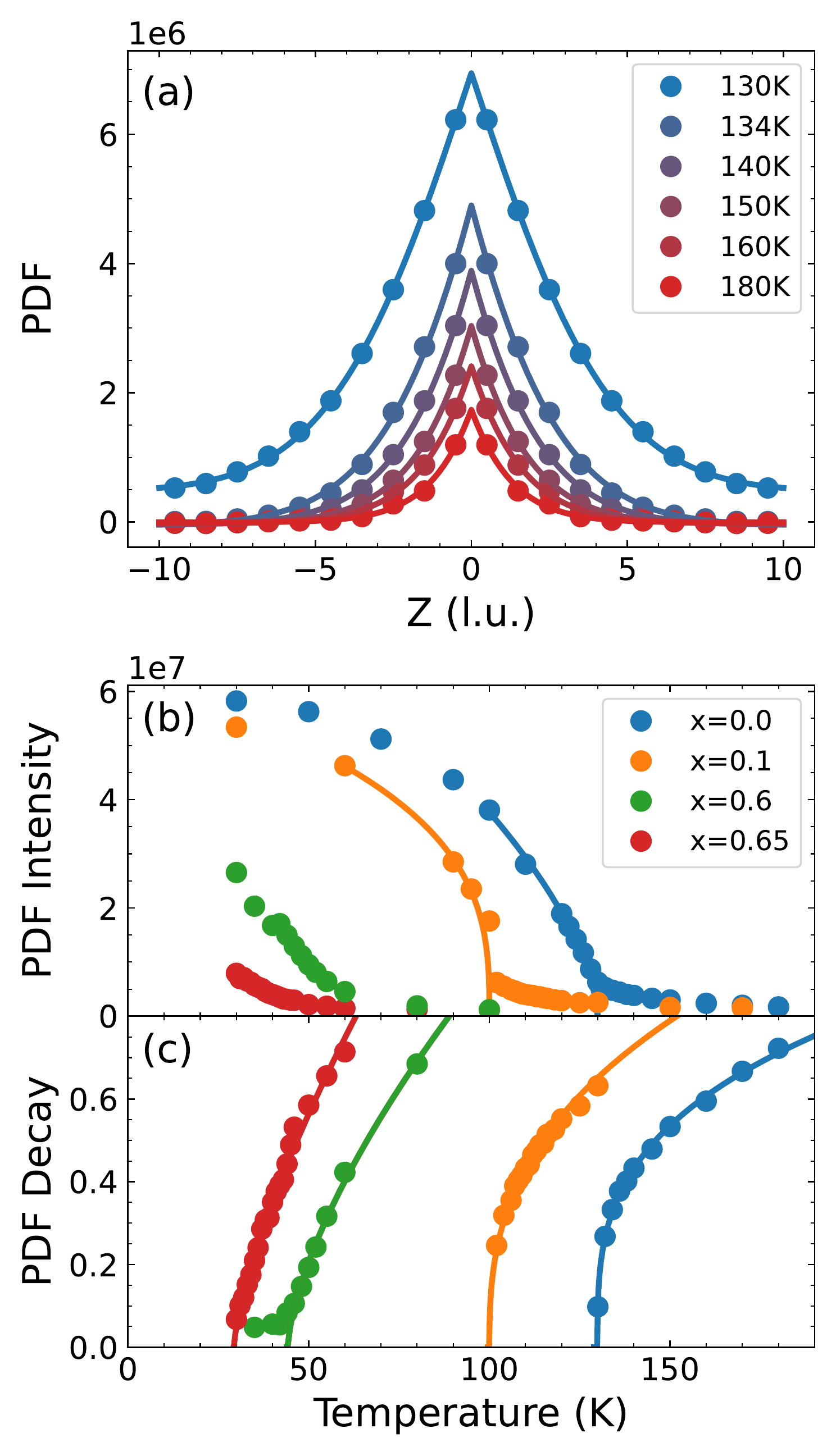}
\caption{\label{fig:fits} (a) Fit of PDF peak intensities along the (0.35,0.65,Z) direction to an exponential decay function in the high-temperature phase at $x=0$. (b) The temperature dependence of the PDF peak intensities extrapolated to the origin at $x=0.0$, 0.1, 0.6, and 0.65  and (c) the inverse correlation lengths determined from the exponential decay constants.}
\end{figure}

Single crystals of \csrs, with $x=0.0$, 0.1, 0.5, 0.6, and 0.65, were synthesized by the Sn flux method \cite{Yang.2010}. Three-dimensional volumes of x-ray scattering were collected on Sector 6-ID-D of the Advanced Photon Source over a range of $\pm15$~\AA$^{-1}$ in all directions by rotating the sample through 360\degree\ using an incident energy of 87.1 keV. The temperature was varied between 30~K and 300~K using a helium/nitrogen cryostream. Further details of the experimental methods are given in the Supplemental Material of Ref. \citenum{Krogstad:2019tc}.

In a previous report, machine learning on the same data sets showed that all the additional superlattice peaks below T$_s$ have a wavevector of ($\frac{1}{2}\frac{1}{2}$0) and its symmetry equivalents with respect to the Bragg peaks \cite{Venderley.2020}, in agreement with previous x-ray diffraction results \cite{Goh:2015cl, Carneiro.2020}. Figure \ref{fig:data}(a,b) shows superlattice peaks in the (HK$\frac{1}{2}$) plane for $x=0$ at 30~K, which weaken in intensity and collapse above T$_s\sim 130$~K. At 150~K, there is evidence of diffuse scattering peaking at superlattice wavevectors from structural fluctuations above T$_s$ (Fig. \ref{fig:data}(c)).

The local distortions were estimated by transforming the data into real space to produce 3D PDFs, both conventional 3D-$\Delta$PDFs, in which the Bragg peaks of the high-temperature phase were removed by using the ``punch-and-fill" method \cite{Weber:2012en,Krogstad:2019tc}, and a modified procedure, in which the data were premultiplied by tiled 3D Gaussian functions centered at superlattice locations. Using these Gaussian windows has the advantage of reducing contributions from thermal diffuse scattering to the PDF, at the cost of an additional convolution with a known waveform. We show in the Supplemental Material (SM) that this does not add artifacts that affect the scientific interpretation of the data.

Figures \ref{fig:data}(d-f) show results for $x=0$ from the modifed 3D PDFs in the Z=0.5 plane. These are maps of the differences in interatomic vector probabilities between the local structure at each temperature and the high-temperature phase. Red (blue) peaks are located at vectors that have higher (lower) probability than the undistorted structure. As expected, the PDF at 30~K shows structural correlations that are long-range, although the intensities are modified due to a Gaussian envelope function from the finite Q-resolution of the original data. At 150~K, however, the correlations only persist over short range, but with similar relative probabilities.

Fits of the $\Delta$PDF peak intensities to an exponential decay as a function of temperature can be used to make estimates of two critical exponents, the order parameter exponent, $\beta$, and the correlation length exponent, $\nu$ (Fig. \ref{fig:fits}). Extrapolations of the $\Delta$PDF intensity to the origin is proportional to the square of the average distortion, which is therefore proportional to the order parameter below T$_s$ apart from a weak contribution from fluctuations close to the transition. Fig. \ref{fig:fits}(b) shows that the average distortion extrapolates to 0 at T$_s$ consistent with a second-order phase transition. The correlation length is determined by the decay constant  above T$_s$ (Fig. \ref{fig:fits}(c)) (see the SM of Ref. \citenum{Krogstad:2019tc} for a more detailed discussion). At $x=0$, $2\beta=0.65$, which is consistent with 3D critical scaling. It falls to 0.35 at $x=0.1$, but the values at higher $x$ are unreliable because T$_s$ was too close to the minimum measuring temperature. We can, however, derive reliable values of $\nu$ over the entire range, with values of 0.31, 0.34, 0.66, and 0.72, at $x=0$, 0.1, 0.6, and 0.65, respectively, \textit{i.e.}, systematically increasing as the quantum critical point is approached.

\begin{figure}[!t]
\includegraphics[width=\columnwidth]{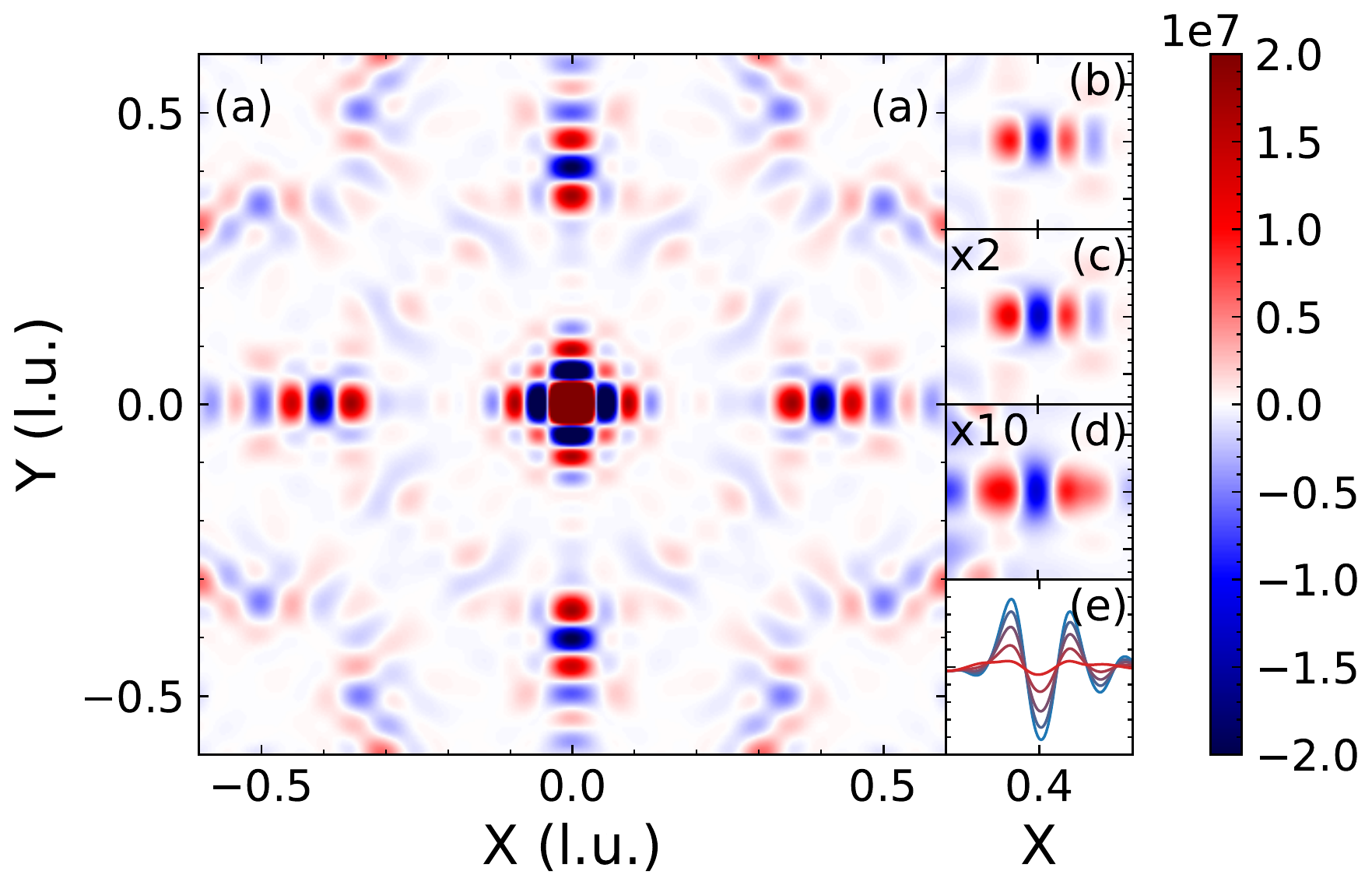} 
\caption{\label{fig:zooms} (a) 3D-PDF map of the Z=0 plane for \srs\ showing the interatomic vectors connecting nearest-neighbor icosahedra, with positive (red) peaks at the distorted positions and negative (blue) peaks at the undistorted positions. These are equivalent to the vectors displayed in Figure 1. X and Y are in lattice units (1 l.u.$=$9.8~\AA). (b-d) An expanded view of the peaks at X$\sim 0.4$ as a function of temperature at (b) 30~K, (c) 100~K, and (d) 150~K. The PDF intensities are scaled by the factors shown in the figures. (e) Line-cuts along the X-axis (Y=Z=0) showing the temperature dependence of the PDF peaks shown in (b-d) at, in order of decreasing intensity, 30~K, 70~K, 100~K, 120~K, and 150~K. An expanded view is shown in the Supplemental Material.}
\end{figure}

Figure \ref{fig:zooms} shows the PDF maps around four interatomic vectors that connect nearest-neighbor icosahedra in the X-Y plane. In the high-temperature phase, there are by symmetry three such vectors parallel to the X, Y, and Z axes, respectively, each of length 0.4 lattice units (1 l.u.$=a=9.8$~\AA), which connect twelve pairs of Sn(2) sites. These split into 48 vectors below T$_s$ based on the crystallographic analysis in Ref. \citenum{Goh:2015cl}, with 12 vectors of length 0.35~l.u. and 12 of length 0.45~l.u.; the remainder stay close to 0.4~l.u. These distortions along the X-axis are illustrated in Fig. \ref{fig:crystal}(b) and the derivation of their magnitude is described in the SM. The  vectors correspond closely to positions of the peaks observed in the 3D-PDF maps (Fig. \ref{fig:zooms}(b))), with the two red peaks arising from distorted vectors and the blue peak at the undistorted vector. Another smaller blue peak occurs at 0.5, but this is the tail of two out-of-plane peaks at (0.5,$\pm 0.05$,$\pm 0.05$).

The side panels of Figure \ref{fig:zooms} shows the positions of the peaks as the temperature increases both below and above T$_s$. In a displacive transition, the three peaks would be expected to merge and cancel each other out at T$_s$, but Fig.  \ref{fig:zooms}(e) shows that there is no temperature variation of the distortion amplitude. Furthermore, the peaks persist in the same locations up to 200~K, 50~K above T$_s$ (see also Fig. S6 in the SM). Within that temperature range, it appears that the vectors connecting nearest-neighbor icosahedra are still distorted even after the longer-range correlations have decayed. Therefore, the PDF analysis provides unambiguous evidence that the amplitude of the local lattice distortions are temperature independent, both within the ordered phase and above. In the SM we show similar results at other compositions.

Our main conclusion is therefore that the structural phase transitions in \csrs\ are of the order-disorder type because the displacement amplitudes are independent of temperature, whereas in displacive phase transitions they should fall to zero at T$_s$ \cite{Cowley.1980}. Our conclusion is in apparent contradiction with the observation of soft phonon modes in inelastic x-ray experiments \cite{Cheung.2017}. As discussed in the introduction, however, it is now generally accepted that few transitions can be rigidly assigned to one or other category on the basis of spectroscopic data above the transition. When the energy barriers between the potential wells are comparable to T$_s$, both soft modes and central peaks may be seen \cite{Bruce:1980bv, Onodera:2004}. It may be significant that the extrapolation of the soft mode energy to zero as a function of temperature falls slightly below T$_s$ (Fig. 3i in Ref. \citenum{Cheung.2017}), as predicted in some order-disorder models \cite{Girshberg:1997, Onodera:2004}, although the difference is small. We also note that the inelastic x-ray measurements show evidence of substantial quasielastic scattering at the superlattice wavevector, consistent with a central peak (Fig. 2 in Ref. \citenum{Cheung.2017}).

Spectroscopic measurements are not therefore inconsistent with the Sn(2) atoms in \csrs\ and related compounds occupying sites located at the minima of multiple potential wells. We believe that it is appropriate to classify the transition as order-disorder, because it is based directly on the temperature-independent amplitude of local Sn(2) displacements, derived from a 3D-PDF analysis that combines diffuse scattering from local fluctuations with superlattice peak intensities, similar to the PDF analysis of polycrystalline materials \cite{Egami.2007}. 

The order-disorder character of the transitions may have implications for the associated electronic transitions in \csrs, and, by extension, related compounds such as \csis. In these compounds, there is substantial evidence of electronic reconstruction at T$_s$, which has been interpreted as resulting from CDW order. There is an increase in the resistivity \cite{Klintberg:2012dt,Goh:2015cl}, as well as a change in sign of the Hall coefficient \cite{Kuo.2014} and a fall in the Knight shift ascribed to a reduction in the density-of-states at the Fermi level \cite{Fang:2014kr,Luo:2018jv}. The $^{119}$Sn NMR spectra are temperature dependent below T$_s$ \cite{Kuo.2014,Kuo:2015cg,Luo:2018jv}, and optical spectroscopy shows evidence of a transfer of spectral weight from a broad Drude component above T$_s$ to a high-frequency peak below that was tentatively identified as the CDW gap energy \cite{Fang:2014kr,Ban.2017}.

For all these reasons, the quasi-skutterudites have been added to the relatively small list of 3D CDW compounds \cite{Klintberg:2012dt}. However, although phonon instabilities are predicted by density functional theory \cite{Tompsett:2014hh}, the nature of the charge modulation has never been established. Charge transfer between Sn(1) and Sn(2) atoms has been proposed based on bond-length arguments \cite{Miraglia.1986,Bordet.1991}; the Sn(1)-Sn(2) bond lengths are all equal above T$_s$ but split into four groups below \cite{Goh:2015cl}. Nevertheless, the thermal ellipsoids are anomalously broad above T$_s$, consistent with our PDF analysis, so it is not clear whether there is a change in charge disproportionation at the transition. 

The persistence of local distortions above the transition therefore calls into question whether these compounds undergo a CDW transition driven by an electronic instability. What then causes the change in electronic properties? One possibility is that the loss of structural phase coherence at T$_s$ produces a concomitant loss of coherence in the electronic states, producing a pseudogap such as seen in the cuprates \cite{Vishik.2018} or intercalated NbSe$_2$ \cite{Chatterjee:2015cw}. This could explain both the broad Drude response above T$_s$ and, if these electronic fluctuations favor superconductivity, the peak in the superconducting dome at the quantum critical point. Photoemission experiments would be valuable to investigate this possibility.

In conclusion, we have shown that, by revealing the amplitude of the local atomic distortions both below and above the transition, 3D-PDF can help to distinguish between order parameters that are determined by a temperature-dependent amplitude (displacive) and those that are determined by the relative site occupations of a number of discrete sites (order-disorder) \cite{Cowley.1980}. In the case of \csrs, our results provide clear evidence of order-disorder character, with implications for the nature of the electronic reconstruction at the phase transition.

\begin{acknowledgments}
We thank Michael Norman for scientific discussions, Johnpierre Paglione and Jeff Lynn for support in the sample preparation and characterization, and Douglas Robinson for technical support during the experiments. This work was supported by the U.S. Department of Energy, Office of Science, Materials Sciences and Engineering Division and used resources of the Advanced Photon Source, a U.S. Department of Energy Office of Science User Facility at Argonne National Laboratory. Crystal structure images were generated using CrystalMaker\textsuperscript{\tiny\textregistered}, CrystalMaker Software Ltd, http://www.crystalmaker.com.

\end{acknowledgments}

\providecommand{\noopsort}[1]{}\providecommand{\singleletter}[1]{#1}%

\end{document}